\documentclass[aps,prd,showpacs,superscriptaddress,nofootinbib,floatfix]{revtex4}
\usepackage{amsmath,amssymb,amsfonts}
\usepackage{hyperref}

\usepackage{graphicx}
\usepackage{color}
\usepackage{epsfig}

\def\be{\begin{equation}}
\def\ee{\end{equation}}
\def\ba{\begin{eqnarray}}
\def\ea{\end{eqnarray}}

\newcommand{\Ham}{H}

\newcommand{\lp}{l_{\text{Pl}}}
\newcommand{\mplanck}{m_{\text{Pl}}}

\begin{document}

\title{Stability of the Schwarzschild Interior in Loop Quantum Gravity}

\author{Christian G.~B\"ohmer}
\email{c.boehmer@ucl.ac.uk}
\affiliation{Department of Mathematics, University College London,
             Gower Street, London, WC1E 6BT, UK}

\author{Kevin Vandersloot}
\email{kevin.vandersloot@port.ac.uk}
\affiliation{Institute of Cosmology \& Gravitation,
             University of Portsmouth, Portsmouth PO1 2EG, UK}

\date{\today}

\begin{abstract}
In recent work, the Schwarzschild interior of a black hole was investigated, incorporating quantum gravitational modifications due to loop quantum gravity. The central Schwarzschild singularity was shown to be replaced by a Nariai type universe. In this brief report we show that this interior solution is stable with respect to small perturbations, in contrast to the classical Nariai universe.
\end{abstract}


\maketitle

In a recent article \cite{Bohmer:2007wi}, the interior of a Schwarzschild black hole was analyzed 
in the framework of loop quantum gravity (LQG -- for reviews see~\cite{ash10,book,Thiemann:2001yy}).
In the analysis it was shown that loop quantum effects can cure the central singularity and
replace it with a Nariai type universe with a Planck scale two-sphere radius. This would seem
to imply that infalling matter would settle in a finite Planckian radius as opposed to collapsing
to the classical singularity, though a more rigorous analysis of the loop effects would be needed
to answer such a question. However, concerns were raised in~\cite{newsci} that the Nariai solution
of classical general relativity is unstable to certain perturbations with the instabilities leading
to a Schwarzschild-de Sitter solution.

In this brief report, we show that the Nariai solution in the interior is stable owing
to the loop quantum effects in contrast to the classical situation. We perform a perturbative
analysis about the Nariai solution using the loop quantum equations of motion. The linearized
perturbation equations can be solved analytically and we show that the perturbations decay if 
evolved towards the center thereby establishing the stability of our solution. Note that the 
issue of stability we investigate is separate from the stability studied in \cite{newBojo}. 
There, the stability of the underlying discrete quantum theory is discussed in the context of 
various quantization ambiguities of the theory. The analysis of \cite{newBojo} is relevant 
to this work in that it supports the choice of $\delta$ parameters that we use in the 
effective Hamiltonian \eqref{d1}.

We briefly recall the equations describing the Schwarzschild interior in the LQG
framework. For more details the reader is referred to \cite{LQGBH, LQGBH2}. 
The interior has been studied in LQG in \cite{na,nb,nc}. For more details, the
reader is referred to those articles. The interior metric is of the form
\be\label{metric}
	ds^2 = -N(t)^2 dt^2 + g_{xx}(t)\, dx^2 + g_{\Omega \Omega}(t)\, d\Omega^2
\ee
with $N(t)$ being the freely specifiable lapse function and
 $d\Omega^2$ representing the unit two-sphere metric. The classical
Schwarzschild interior solution is given by 
\be \label{Swcl}
	N(t)^2 = \bigg(\frac{2m}{t}-1\bigg)^{-1}, \quad g_{xx}(t) = \bigg(\frac{2m}{t}-1\bigg), \quad g_{\Omega \Omega}(t) = t^2
\ee
for $t$ in the range $t \in [0, 2m]$, where $m$ is the mass of the black hole. 
Here, $t=0$ corresponds to the classical singularity and $t=2m$ being the 
horizon. Loop quantum gravity uses a separate set of variables than the metric ones. 
The metric components are encoded in triad components $p_b, p_c$ which are related as
\ba
	g_{xx} = \frac{p_b^2}{|p_c|}, \quad
	g_{\Omega \Omega} = |p_c|   \,.
\ea
Note that the physical radius of the two-sphere is determined from the $g_{\Omega \Omega}$ or
$p_c$ components. In particular, the classical singularity corresponds to a vanishing
two-sphere radius where $g_{\Omega \Omega} = p_c = 0$. The LQG variables also consist of
connection components $b, c$ which are conjugate to the triad components. We focus on
the triad components for interpretation since they are directly related to the metric components.

The Nariai solution~\cite{Nariai2,Nariai1} is a spherically symmetric solution of the
classical field equations of general relativity where the two-sphere radius is assumed to be
constant and the model is sourced by a positive cosmological constant $\Lambda$ with the possible
inclusion of charge. For the case of no charge, the metric is given
by
\be\label{lam11}
      ds^2 = -dt^2 + \frac{1}{\Lambda} \cosh^2(\sqrt{\Lambda}t) dx^2 +
\frac{1}{\Lambda} d\Omega^2      
\ee
the form of which is equivalent to \eqref{metric}.
The triad components are thus given by
\be \label{pNariai}
	p_b = \frac{1}{\Lambda} \cosh(\sqrt{\Lambda}t), \quad  p_c = \frac{1}{\Lambda} \,.
\ee
The classical Nariai solutions are quantum mechanically unstable. 
Such instabilities lead to perturbations in the two-sphere radius
$p_c$ and the solution decays into a Schwarzschild-de Sitter black hole
solution (or Reissner-Nordstrom-de Sitter for the charged case)~\cite{Ginsparg:1982rs,Bousso:1996au,Bousso:1996pn}.

The loop quantum effects are derived from a modified effective Hamiltonian 
\cite{Bohmer:2007wi,Modesto:2006mx} based on the phase-space variables $b, c, p_b, p_c$ 
which is given by (note that we use the improved quantum Hamiltonian of 
section IIIB in \cite{Bohmer:2007wi})
\begin{align} 
      \Ham_{\rm eff} = -\frac{N}{2 G \gamma^2} \biggl[
      2 \frac{\sin b \delta_b}{\delta_b} \frac{\sin \delta_c c}{\delta_c} \sqrt{p_c} +
      \Bigl(\frac{\sin^2\negmedspace b \delta_b}{\delta_b^2} + \gamma^2\Bigr)
      \frac{p_b}{\sqrt{p_c}} \biggr].
      \label{d1}
\end{align}
Here $\gamma$ is known as the Barbero-Immirzi parameter, a real valued ambiguity parameter
of LQG than can be constrained from black-hole entropy considerations. The parameters
$\delta_b, \delta_c$ are functions of the triad variables, the exact form
given in~\cite{Bohmer:2007wi}. These parameters directly measure the magnitude of
the loop quantum effects. In the limit of $\delta_b, \delta_c \rightarrow 0$, the
classical behavior is recovered.

Choosing the lapse function to be $N=\gamma \sqrt{p_c} \delta_b/(\sin b \delta_b)$, the complete set of equations 
derived from the effective Hamiltonian \eqref{d1} are given by
\begin{eqnarray}
  \dot{c} &=&
  -\frac{\sin c \delta_c}{\delta_c} -\frac{1}{2}\frac{\sin\delta_b b}{\delta_c}
  -c \cos c \delta_c \nonumber \\
  &&+ \frac{1}{2}\frac{\delta_b}{\delta_c} b
  \cos b \delta_b
  \Bigl(1-\frac{\gamma^2 \delta_b^2}{\sin^2\negmedspace b \delta_b}\Bigr)
  +\frac{\gamma^2}{2}\frac{\delta_b^2}{\delta_c}\frac{1}{\sin b \delta_b}
  \label{d3}\\
  \dot{p_c} &=& 2 p_c \cos c \delta_c
  \label{d4}\\
  \dot{b} &=&
  -\frac{1}{2}\Bigl(\frac{\sin  b \delta_b}{\delta_b} +
  \frac{\gamma^2 \delta_b}{\sin b \delta_b}\Bigr) -
  \frac{\sin c \delta_c}{\delta_b}
  + \frac{\delta_c}{\delta_b} c \cos c \delta_c
  \label{d5}\\
  \dot{p_b} &=&
  \frac{1}{2}\cos b \delta_b \Bigl( 1 - 
  \frac{\gamma^2 \delta_b^2}{\sin^2\negmedspace b \delta_b}\Bigr) p_b.
  \label{d6}
\end{eqnarray}
With this choice of lapse, the classical singularity would correspond to 
$T \rightarrow -\infty$. Furthermore, the system is a constrained type whereby
the Hamiltonian should vanish, providing an additional equation of motion.

\begin{figure}[!ht]
\noindent
\begin{minipage}[h]{.48\linewidth}
\centering\epsfig{figure=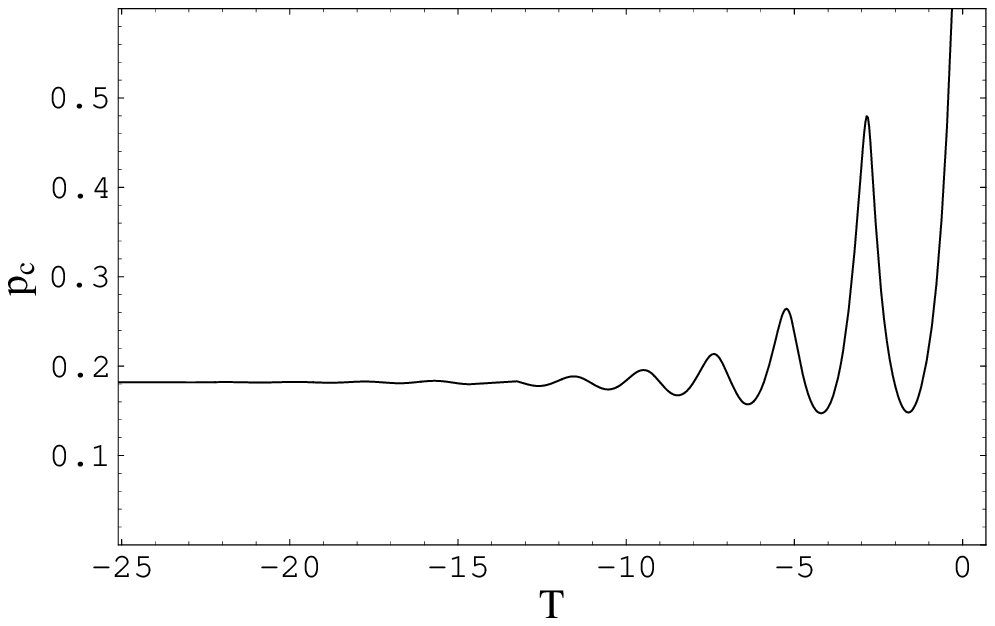,width=\linewidth}
\end{minipage}\hfill
\begin{minipage}[h]{.48\linewidth}
\centering\epsfig{figure=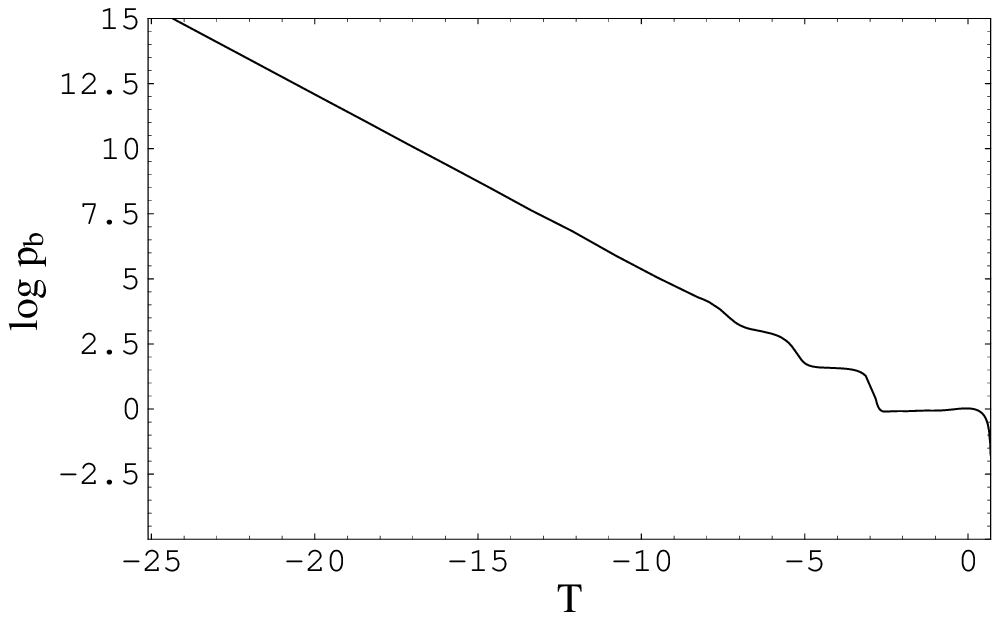,width=\linewidth}
\end{minipage}
\caption{In the left figure $p_c(T)$ is plotted which tends towards a constant as $T\rightarrow -\infty$. The right figure shows $p_b(T)$ which grows like $\exp(-\alpha T)$ asymptotically. This asymptotic behavior is indicative of a Nariai type metric.}
\label{fig:deltawell2}
\end{figure}

The results of a numerical solution to the equations of motion are shown in figures
\ref{fig:deltawell2}. The results indicate that $p_c$ tends to a constant
value as $T$ tends to minus infinity while $p_b$ grows as an exponential as in
the Nariai case, see Eqs.~\eqref{pNariai}. 
This is indicative of the space-time metric approaching a
Nariai type universe as the classical singularity is approached.

In the limit of $T \rightarrow -\infty$, there is an exact asymptotic solution
to the quantum equations of motion to which the solution shown in
figure \ref{fig:deltawell2} tends to.
This solution is characterized by
\begin{alignat}{2}
  b &= \bar{b}, & \qquad p_b &= \bar{p}_b\, e^{-\alpha T},
  \nonumber \\
  c &= \bar{c}\, e^{-\alpha T}, & \qquad p_c &= \bar{p}_c,
  \label{d7}
\end{alignat}
where the barred quantities and $\alpha$ are constants. It is clear that this
asymptotic solution corresponds to a Nariai type solution \eqref{pNariai} in
the large negative $T$ limit. From the equations of motion we can determine 
the asymptotic values of the constants $\bar{p}_c, \bar{b}, \alpha,$ and 
$\bar{c}/\bar{p}_b$. A root find algorithm can be used to determine the values 
of the constants which are given by \cite{Bohmer:2007wi}
\begin{alignat}{3}
  \bar{b} &\approx 0.156, & \qquad \bar{p}_c &\approx 0.182\, \lp^2, \qquad \alpha &\approx 0.670
  \nonumber \\
  \bar{c}/\bar{p}_b &\approx -2.290 \, \mplanck^2, &\qquad \bar{N} &\approx 0.689, &
  \label{d13}
\end{alignat}
where $\bar{N}$ is the asymptotic value of the lapse which also behaves as a constant.
 It is noteworthy that the fixed two-sphere radius is Planckian in length
since the value of $p_c$ approaches a constant Planckian value of $0.182 \lp^2$.
Furthermore this value is independent of the mass of the black hole. 

\begin{figure}[!ht]
\noindent
\begin{minipage}[h]{.48\linewidth}
\centering\epsfig{figure=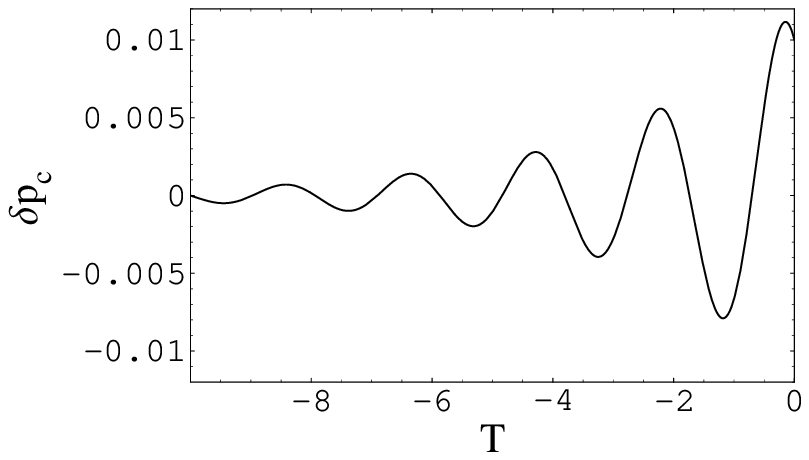,width=\linewidth}
\end{minipage}\hfill
\begin{minipage}[h]{.48\linewidth}
\centering\epsfig{figure=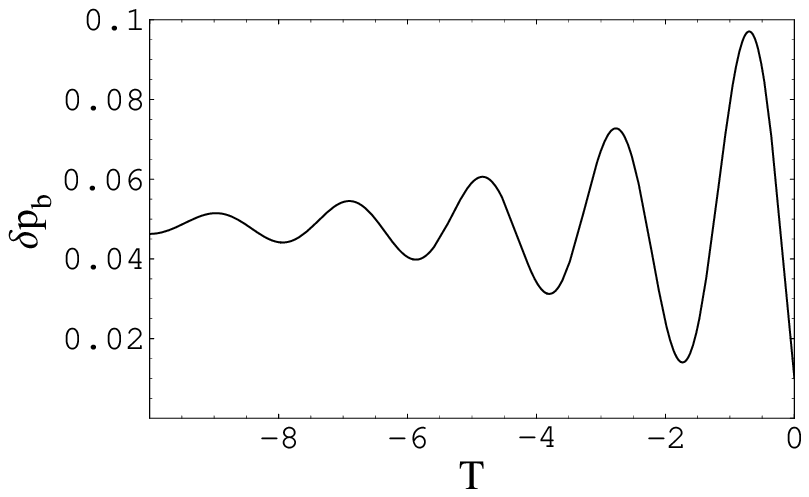,width=\linewidth}
\end{minipage}
\caption{In the left figure we plotted the perturbation $\delta p_c$ as a function of $T$. We see that the amplitude of the initial perturbation $\delta=0.01$ decreases towards the location of the classical singularity and eventually vanishes. The right figure show the behavior of the perturbation $\delta c$ which also shows a damped oscillation.}
\label{fig:pert}
\end{figure}

Given that the classical Nariai solution is unstable to perturbations of the two-sphere
radius, we can now ask the same question in the context of the loop quantum equations.
We consider perturbations around the asymptotic solution \eqref{d7} of
the form
\begin{align}
  c &= \bar{c} e^{-\alpha T} (1 + \varepsilon c^{(1)}(T)),\\
  p_c &= \bar{p}_c + \varepsilon p_c^{(1)}(T),\\
  b &= \bar{b} + \varepsilon b^{(1)}(T),\\
  p_b &= \bar{p}_b e^{-\alpha T} (1 + \varepsilon p_b^{(1)}(T)).
\end{align}
Now, we insert these expressions into the equations of motion and linearize the equations with respect to $\varepsilon$. Since we perturb about the Nariai background the zero order equations are identically satisfied. Since the four equations of motion are constrained
due to the vanishing of the Hamiltonian, we are effectively left with three independent equations. This linear system of differential equations describing the perturbations then becomes
\begin{align}
  \frac{d}{dT}
  \begin{pmatrix} c^{(1)} \\ p_c^{(1)} \\ p_b^{(1)} \end{pmatrix} =
  \begin{pmatrix}
  -2.2408 & -6.4778 & 2.2408 \\
  0.5717 & 1.5708 & -0.5717 \\
  0 & 16.0296 & 0
  \end{pmatrix}
  \begin{pmatrix} c^{(1)} \\ p_c^{(1)} \\ p_b^{(1)} \end{pmatrix},
\end{align}
and is solved by
\begin{align}
  c^{(1)} &= (4.8426 - 3.8426 e^{\delta T}\cos(\omega T) + 
  2.5552  e^{\delta T}\sin(\omega T))c^{(1)}_0)\xi,\\
  p_c^{(1)} &= (\cos(\omega T) - 0.6271 \sin(\omega T) )e^{\delta T}\xi,\\
  p_b^{(1)} &= (4.8426 - 3.8426 e^{\delta T}\cos(\omega T) - 
  4.8510 e^{\delta T}\sin(\omega T))c^{(1)}_0)\xi,
\end{align}
where $\omega=3.0390$, $\delta=0.3350$ and where $\xi$ denotes the initial perturbation. Note that we are interested in the dynamical behavior of the perturbation close to the classical singularity, located at $T=-\infty$. Therefore, one can immediately conclude that the perturbations damp off. A plot of the perturbation behavior is shown in figure \ref{fig:pert}, showing the damped
oscillatory behavior for $T\rightarrow -\infty$.

We have thus shown that the Nariai behavior near the classical singularity
is a stable solution. The quantum interior will not decay into a separate 
Schwarzschild-de Sitter universe in contrast to the the classical Nariai 
universe. However, the most important issue of how the loop modifications 
affect a real inhomogeneous collapse scenario remains an open problem.

\acknowledgments 
KV is supported by the Marie Curie Incoming International Grant M1F1-CT-2006-022239.

\end{document}